\documentclass[journal=iecred,manuscript=article]{achemso} 
\usepackage[T1]{fontenc} \usepackage{amsmath}
\usepackage{gensymb}

\usepackage{pdfcomment} \pdfcommentsetup{color={1 1 0}, fontcolor={0 0 0}}

\newcommand{\dd}{\,\mathrm{d}}
\DeclareMathOperator\erfc{erfc}
\newcommand{\supp}{Supporting Information}
\newcommand{\DINF}[2][]{\ensuremath{D^{\infty}_{#2\mathrm{#1}}}}
\newcommand{\DIJ}[2][]{\ensuremath{D_{#2\mathrm{#1}}}}
\newcommand{\cotwo}[1][]{                          \ifmmode
                            ^{#1}\mathrm{CO}_{2}                          \else
                            \textsuperscript{#1}CO\textsubscript{2}                          \fi}
\newcommand{\DcotwoSup}[2][]{\ensuremath{\DIJ[]{\cotwo[]}^{#2\mathrm{#1}}}}

\usepackage{xcolor}
\newcommand{\changed}[1]{#1}
\usepackage{amssymb}

\usepackage{xr-hyper}         \makeatletter
    \newcommand*{\addFileDependency}[1]{      \typeout{(#1)}
      \@addtofilelist{#1}
      \IfFileExists{#1}{}{\typeout{No file #1.}}
    }
    \makeatother
    
    \newcommand*{\myexternaldocument}[1]{        \externaldocument{#1}        \addFileDependency{#1.tex}        \addFileDependency{#1.aux}    }
    
    \myexternaldocument{SI}
    
\author{Oliver~Großmann}
\author{Sarah~Mross}
\author{Jens~Wagner}
\author{Kerstin~Münnemann}
\author{Fabian~Jirasek}
\author{Hans~Hasse}
\email{hans.hasse@rptu.de}            
\affiliation{Laboratory of Engineering Thermodynamics (LTD), RPTU Kaiserslautern, Erwin-Schrödinger-Straße 44, 67663, Kaiserslautern, Germany}

\title{Measurements of Diffusion Coefficients of \cotwo[] in 1-Butanol with a New Laminar Jet Apparatus and PFG-NMR, Pointing to Interfacial Mass Transfer Effects}

\begin{document}

    \begin{tocentry}
        \includegraphics[width = 8.25cm, height = 4.45 cm]{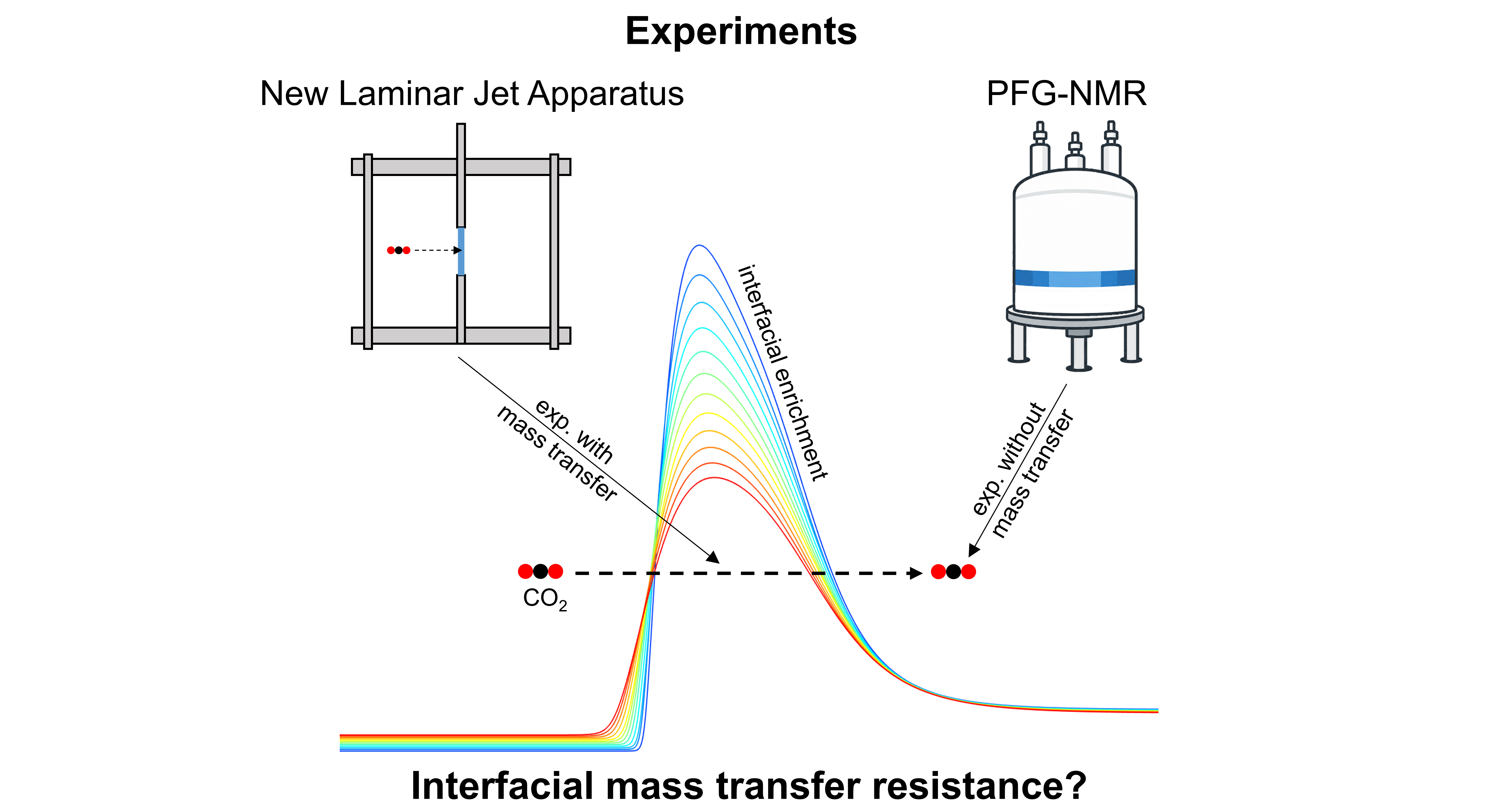}
    \end{tocentry}
    
    \begin{abstract}
                                                        \changed{A novel laminar jet apparatus (LJA) was constructed for precise gas–liquid mass transfer measurements up to 12~bar, significantly extending the technique's operating window. It was applied to carbon dioxide + 1-butanol between 283~K and 333~K. Corresponding measurements were performed using pulsed field gradient NMR spectroscopy (PFG-NMR), which does not involve interfacial mass transfer. Fick diffusion coefficients from LJA and self-diffusion coefficients from PFG-NMR were compared in the limit of infinite dilution, where both must coincide. Both methods show consistent trends, but LJA data are systematically lower. Experimental errors cannot explain the deviations. We therefore hypothesize an additional gas–liquid interfacial mass transfer resistance. PCP-SAFT combined with density gradient theory predicts high \cotwo[] enrichment at the interface, and the deviations correlate with this enrichment. Whether such an interfacial resistance exists and is caused by enrichment remains to be established in future studies.}        
    \end{abstract}
    
    \newpage
        
    \section{Introduction}

        Gas-liquid mass transfer plays a central role in industrial reaction and separations processes as well as in many processes in nature. Common theories of mass transfer consider mass transfer resistances on  both sides of the interface, assuming that phase equilibrium is established at the interface. This assumption has been challenged and there is a long-standing debate in the literature whether the enrichment of a diffusing species $i$ at the interface could lead to an additional mass transfer resistance (for more details, see below).

        This hypothesis of an additional mass transfer resistance at the interface can, in principle, be tested in laboratory experiments with a laminar jet apparatus (LJA). In the LJA, a gas is absorbed by a liquid laminar jet. For the laminar jet, the liquid-side mass transfer resistance is known a priori as a function of the diffusion coefficient of the gas in the liquid.
        Therefore, assuming that the gas-side mass transfer is negligible, the \changed{diffusion coefficient} can be calculated from the measured \changed{mass transfer rate} and vice versa. If the \changed{diffusion coefficient} is known from a reliable reference experiment in the homogeneous phase, the calculated mass transfer rate can be compared to experimental results from the LJA. 
        If the experimental values are found to be consistently lower than expected, the hypothesis of an additional mass transfer resistance at the interface is supported. We have carried out a corresponding test in the present work.

        In order to be able to test the hypothesis, we first had to design a suitable LJA. The LJA is often used for investigations of absorption of gases into chemically reactive solvents at ambient pressure, for which the mass transfer rates are high\cite{matsuyama_research_1950, manogue_kinetics_1960, aboudheir_kinetics_2003, ramachandran_kinetics_2006}. However, to put the hypothesis to a test, simple physical solvents are preferred, which requires an increase of the system pressure to achieve mass transfer rates that enable a reliable evaluation.         Hence, the LJA must be operated at elevated pressures. As such an apparatus did not exist and has also not been described in the literature, we had to construct it and set it up.

        Besides the apparatus, also the availability of suitable reference data for the \changed{diffusion coefficient} was an issue in the project. To test the hypothesis, the reference data should stem from an experimental method that does not involve a gas-liquid interface. However, such data are rare in the literature, especially when high quality is required. We have, therefore, decided to carry out own reference measurements. For these measurements, pulsed field gradient nuclear magnetic resonance spectroscopy (PFG-NMR) was used, a technique in which the \changed{diffusion coefficient} is determined from experiments on equilibrated solutions. For background information on experiments with the LJA and with PFG-NMR, we refer the reader to the literature\cite{matsuyama_research_1950, scriven_fluid_1959, duda_laminar_1968, danckwerts_gas-liquid_1970, brignole_mass_1981, weingartner_nmr_2002, bellaire_pfg-nmr_2020, Bellaire2022, mross_diffusion_2024, phuong_determination_2024, grossmann_hyrodynamics}.

        Rather than comparing predicted and measured mass transfer rates, we have compared the values for the \changed{diffusion coefficient} obtained with the LJA and by PFG-NMR. The hypothesis would be supported if the diffusion coefficients from the LJA were consistently lower than those from PFG-NMR. 
                                                                                \changed{In comparing results from these two experimental methods, it must be considered that the LJA yields the Fick diffusion coefficient, whereas PFG-NMR yields the self-diffusion coefficient. However, these diffusion coefficients coincide in the limit of infinite dilution of the solute, so that a direct comparison is possible in this limit. To keep the notation brief, we therefore use the symbol \DIJ[]{\cotwo[]} throughout for the diffusion coefficient of carbon dioxide in 1-butanol and do not explicitly distinguish between self-diffusion and Fick diffusion when the meaning is clear from the context.}

        The topic of interfacial resistance in gas-liquid mass transfer, specifically in LJAs, has been raised in the literature since the 1950s. While several authors have argued in favor of such an effect (e.g. \citet{chiang_interfacial_1959,gupta_effect_1984}), others have challenged that notion and proposed modeling inaccuracies (e.g. in the surface velocity) \cite{raimondi_interfacial_1959,fosberg_interphase_1967} or contamination with surfactants \cite{scriven_phase_1958,duda_laminar_1968} and other effects as explanations of an observed apparent resistance. 
        The accurate description of gas-liquid mass transfer remains a subject of ongoing discussion \cite{huthwelker_analytical_1996,ma_note_2005} and was recently revived by computational studies on interfacial properties of binary fluid mixtures, in which a so-called enrichment was observed, which is related to the preferential adsorption of one of the mixture's components to the fluid interface \cite{becker_interfacial_2016}. 
        This effect has been predicted by several independent methods -- namely molecular dynamics simulations, density gradient theory (DGT), and density functional theory (DFT) -- in mixtures of model fluids and real fluid mixtures alike \cite{enders_interfacial_2008, klink_density_2014, stephan_vaporliquid_2018, stephan_interfacial_2019, staubach_interfacial_2022, stephan_vaporliquid_2024}. 
        Enrichment is especially pronounced in gas-liquid mixtures with a wide-boiling VLE \cite{stephan_molecular_2020, stephan_enrichment_2020}, such as those typically encountered in absorption. Under favorable conditions, namely low temperature and high dilution of the component $i$, the local concentration of that component at the interface may be several times higher than either of its bulk phase concentrations\cite{stephan_enrichment_2020}.
        Unfortunately, the enrichment cannot be studied directly by experiments, due to the nanoscale nature of the phenomenon. 
        There is currently a discussion on the macroscopic effects of enrichment on the transport through vapor-liquid interfaces\cite{nagl_interfacial_2020, klink_density_2014, garrido_understanding_2015, schaefer_mass_2023, braten_molecular_2023, grossmann_vapor-liquid_2024}, in which a hypothesis is that a high enrichment at the interface may cause an additional mass transfer resistance. 
        The test system we have studied was carbon dioxide (\cotwo[]) + 1-butanol (1-BuOH). It was selected for two reasons: Firstly, because 1-BuOH is a good solvent for \cotwo[], leading to fairly high mass transfer rates, which enables studies over a wide range of pressures in the LJA, and, secondly, because based on previous experience, it was expected to exhibit a high enrichment of \cotwo[] at the interface.
        Deviations between the two measured data sets were compared to results for the enrichment in the studied system, which were calculated using the PCP-SAFT equation of state (EOS)\cite{gross_perturbed-chain_2001, gross_application_2002, gross_equation--state_2005, gross_equation--state_2006, vrabec_vapor-liquid_2008} and DGT\cite{rowlinson_molecular_1982, cornelisse_fundamentals_1997, miqueu_modelling_2003}.

        \changed{Only three experimental data points are available in the literature for the studied system, all of them at 298~K\cite{hikita_liquid_1959, takeuchi_simultaneous_1975}. Therefore, also literature data for other systems of the type \cotwo[] + 1-alkanol were included in the discussion of the deviations.}

        This paper is structured as follows. First, the new LJA apparatus and its operation are described, followed by the presentation of the model used to determine \DIJ[]{\cotwo[]} from the LJA measurements. Subsequently, the PFG-NMR reference measurements are introduced and the calculation of the enrichment in the studied system is explained. The experimental results obtained from both methods are then presented and critically compared. The observed differences are then assessed in light of the computational results for the enrichment in the studied system, and conclusions are drawn.

    \section{Methods}

        \subsection{Laminar Jet Apparatus}
															
        A novel pressurized LJA was designed specifically for this work.         Significant effort was dedicated to ensuring steady jet operation and precise measurement of all process parameters.
        The following section begins with a detailed technical description of the LJA construction. Next, the experimental procedure is described and raw measurement data are shown for an exemplary set of conditions. Lastly, the modeling approach and equations used for the evaluation of the experiments are presented.

        \subsubsection{Jet Chamber}

        The centerpiece of the LJA is the jet chamber. It is a hermetically sealed vessel in which the absorption from gas into the liquid jet takes place.

\begin{figure}
            \centering
            \includegraphics[height=0.8\textheight]{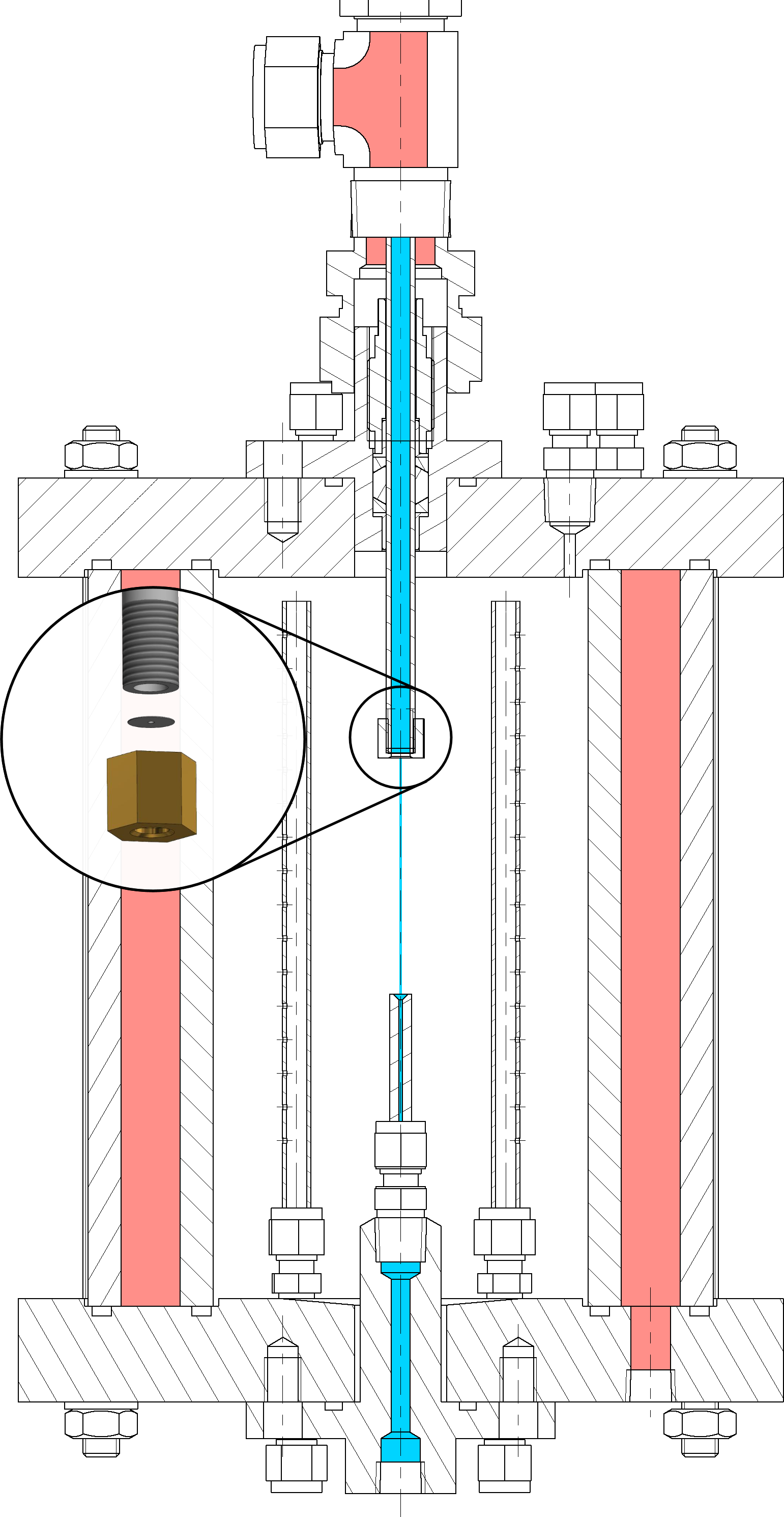}
            \caption{Technical drawing of the jet chamber with zoom on the orifice plate. The parts carrying the process liquid are filled in blue color; the parts with thermostatting fluid are filled in red color.}
            \label{fig:zelle_skizze}
        \end{figure}

        A technical drawing of the chamber is shown in Figure~\ref{fig:zelle_skizze}. It is designed as a double jacketed glass vessel with stainless steel flanges on the top and bottom. The inner and outer glass jackets (9~mm thickness, 120~mm height, 120~mm and 170~mm outer diameter, respectively) are fabricated from flame-polished borosilicate 3.3 glass (for low thermal expansion coefficients). The stainless steel flanges (208~mm diameter, 30~mm height) are joined by threaded rods. There are through holes in the flanges that allow for the connection of pressure and temperature sensors, a relief valve, and the temperature control medium (red in Figure~\ref{fig:zelle_skizze}), as well as four gas distributors in the form of vertical pipes with evenly-spaced radial holes. The inner volume of the chamber is approximately 1.6~dm$^3$.
        
        The process liquid (blue in Figure~\ref{fig:zelle_skizze}) is introduced into the chamber via a capillary tube and exits the chamber through a plexiglas capture funnel. The capillary tube (650~mm length, 5~mm inner diameter) can be adjusted vertically using a spindle lifting drive, enabling jet lengths from 0 to over 200~mm with a precision of about $\pm$0.01~mm. The capillary tube terminates in a square-edge orifice plate (0.1~mm thickness, 1~mm hole diameter, held by a union nut, cf. inlet of Figure~\ref{fig:zelle_skizze}), which serves as the nozzle. A jet of 0.5~mm radius with approximately uniform velocity profile is generated by that nozzle. The fluid dynamics of jets generated by such nozzles have recently been studied in detail\cite{grossmann_hyrodynamics}. 
        
        The capture funnel is mounted to the capture flange via a 1/4" threaded pipe.
        The capture flange can be adjusted perpendicular to the jet, allowing precise alignment of the capture funnel with the jet (technical details are given in the \supp{}).

        The setup was designed and tested for pressures up to 15~bar and temperatures from 10~\degree{}C to 80~\degree{}C. 
        \subsubsection{Periphery}

\begin{figure}
            \centering
            \includegraphics[width=0.85\textwidth]{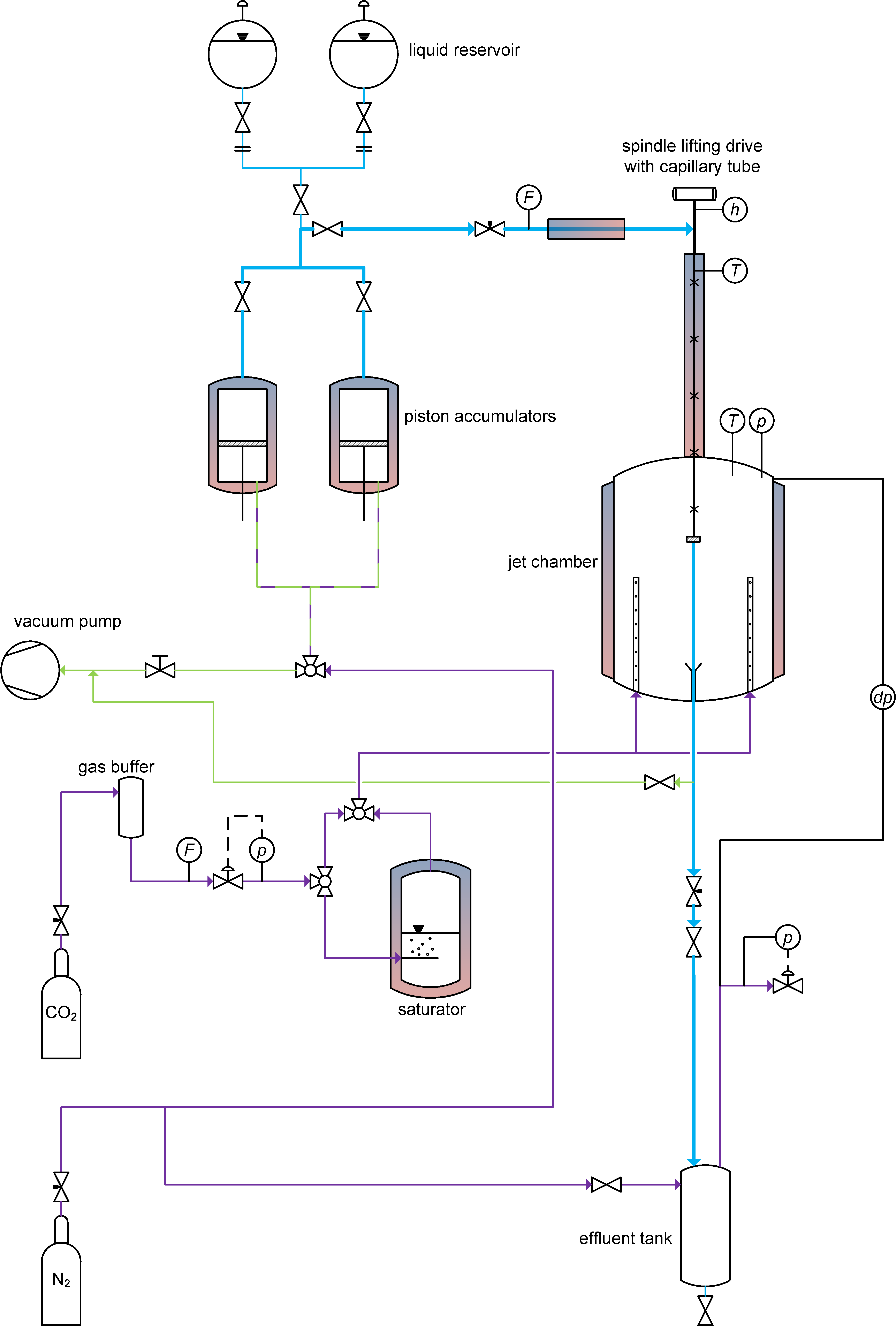}
            \caption{Simplified process flow diagram of the laminar jet apparatus, showing only the valves and measurements relevant for measurement operation. Piping colors indicate:  blue: liquid, purple: gas, green: vacuum.}
            \label{fig:fliessbild}
        \end{figure}

       The periphery of the jet chamber includes a fluid delivery system, the process gas feed, a thermostatting system, and the effluent capture. Figure~\ref{fig:fliessbild} shows a simplified flowsheet of the LJA; a complete diagram is provided in the \supp{}.

              Fluid delivery was provided by pneumatically pressurized piston accumulators, which supplied a nearly pulsation-free liquid flow, ensuring stable jet operation.
              Two custom-built piston accumulators with an internal volume of about 4~dm$^3$ each were used. They can be operated in parallel or sequentially. The gas side connects to either a vacuum pump (fluid intake) or a nitrogen tank (fluid delivery); the liquid side is either connected to the reservoir containing degassed liquid (fluid intake) or via a mass flow meter (Bronkhorst CORI-FLOW M54, accuracy $\pm0.2\%$) and a needle valve for flow regulation to the capillary tube of the jet chamber (fluid delivery).

        The process gas is drawn from a gas cylinder via a buffer tank. The gas flow was         measured (Bronkhorst EL-FLOW SELECT F-111B, accuracy         $\pm0.6\%$) and controlled by a pressure regulator (Bronkhorst EL-PRESS P-602CV) to hold the chamber setpoint pressure. The gas is then fed either through a saturator or         directly into the four distributors of the jet chamber.

        Temperatures are controlled by two thermostatting loops; one comprising the piston accumulators and feed line, and the other comprising the saturator and the jet chamber. Each loop is serviced by a thermostat (Julabo F32-HE) and equipped with temperature sensors. Furthermore, the temperature is measured in the liquid feed at the capillary inlet and in the gas phase of the jet chamber by Pt100 resistance thermometers.

        The effluent tank must be held at a pressure slightly below the chamber pressure to avoid gas breakthrough through the funnel.
        It was pressurized with nitrogen and the pressure was adjusted by a control valve.
                The liquid flow from the chamber to the tank is manually adjusted via a needle valve to maintain a constant liquid level in the capture funnel. The volume of the stainless steel effluent tank is 10~dm$^3$ to accommodate the entire liquid feed; 
        the dimensioning of the present LJA allows for about 1~h of continuous jet operation.        
        \subsubsection{Experiments}    

        The process liquid -- here 1-butanol (1-BuOH, $\ge$0.995~g/g, Merck KGaA) -- is filled into two glass reservoirs of about 4~dm$^3$ each. The liquid is then degassed, first by flushing with helium gas, then by vacuum. The reservoirs are connected to the LJA and the liquid is drawn into the thermostatted piston accumulators by applying a vacuum to the gas side of the pistons.

        The jet chamber is sealed hermetically, evacuated and thermostatted. Then, process gas -- here carbon dioxide (CO$_{2}$, $\ge$0.99995~l/l, Air Liquide N45) -- is added to the chamber until the experiment set point pressure is reached. Next, the effluent tank is pressurized with inert gas -- here nitrogen (N$_{2}$, Air Liquide ALPHAGAZ~1) -- to a setpoint slightly below the chamber pressure. We found a differential pressure of about $-300$~mbar to provide the best operability.

        When all temperatures and pressures are in steady-state, the experiment is started. 
        For this, the gas side of the piston accumulators is switched to N$_{2}$ and the valves connecting the liquid side to the jet chamber are opened. Immediately afterwards, right as the liquid jet is forming, the valve to the effluent tank is opened. The control valves in the liquid lines before and after the jet chamber are adjusted; the former to the desired flow rate and the latter to keep the liquid level in the capture funnel at a constant position near the tip.

        During the experiment, the jet length is repeatedly increased in small increments -- here in 3~mm steps. The maximum jet length depends on the hydrodynamics, e.g., the breakup length, see \cite{grossmann_hyrodynamics}. 
                For each jet length, a period of approximately 2--5~min is required for the system to reach and maintain a pseudo steady state, in which the make-up gas flow is constant and equals the mass transfer rate in the jet. It should, however, be noted that maintaining a constant level in the funnel usually requires frequent adjustment of the control valve in the effluent line, and also the gas flow rate fluctuates (due to the flow controller behavior). Hence, the results for the pseudo steady state are averaged for each jet length to obtain the data basis for the subsequent evaluation.

\begin{figure}
            \makebox[\textwidth][c]{
                \includegraphics[width=1.2\textwidth]{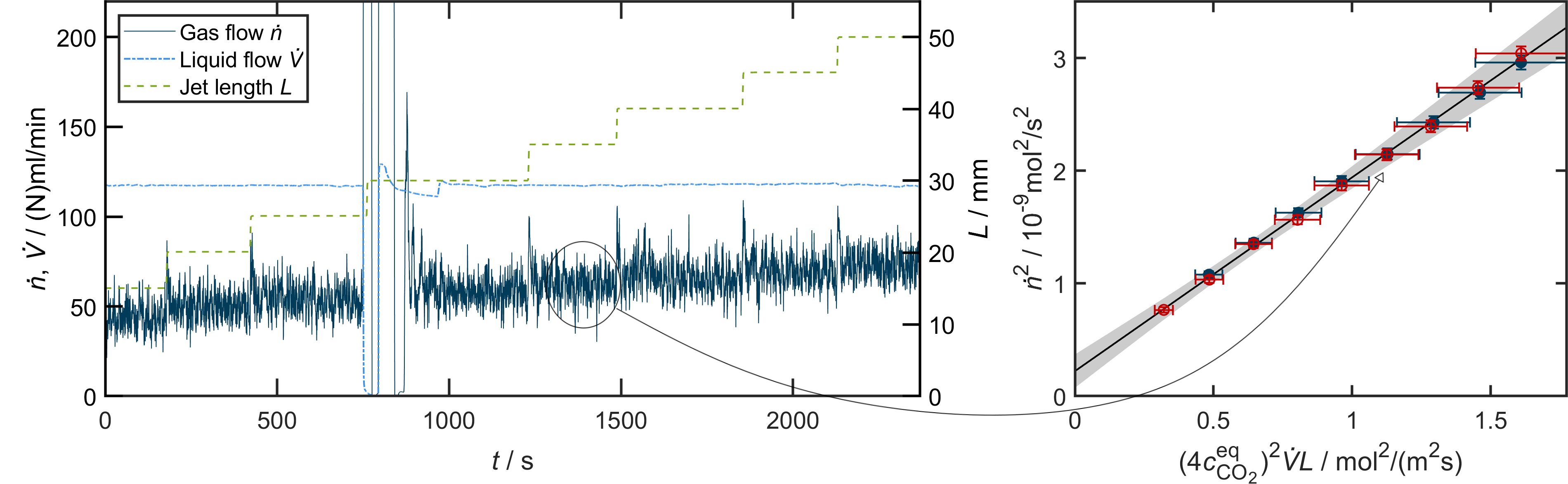}}
            \caption{            Example for results from an LJA experiment (data for 283.15~K and 9~bar). The left panel shows primary data for the \cotwo[] flow rate $\dot{n}_{\cotwo[]}$ and the liquid flow rate $\dot{V}$ as a function of time $t$. The jet length $L$ is also indicated, which is varied in the experiment. The large disturbances after about $t=750$~s are caused by switching between the piston accumulators that supply the liquid. The dark blue symbols in the right panel represent the averaged experimental results for the different pseudo steady state periods. The errorbars are the measurement uncertainties. The data set in red stems from an independent repeat experiment and confirms the excellent repeatability. The line is a linear fit to both datasets based on Eq.~(\ref{eqn:evaluationEquationNdot}) and the shaded area is the 95\% confidence interval of the prediction. The value of the diffusion coefficient is found from the slope of the line.}            \label{fig:combinedExp}
        \end{figure}

                Primary experimental results from a typical LJA experiment, i.e., the gas flow of \cotwo[] (in Nml/min, as obtained from the sensor) as a function of time $t$ for different jet lengths, are shown in the left panel of Figure~\ref{fig:combinedExp}. The experiment was carried out at 283.15~K and 9~bar; the absorption rate varies between 45 and 75~Nml/min for jet lengths between 15 and 50~mm.
                The plot is divided into two sections (at about 750~s) by the switch from piston \#1 to piston \#2, during which the liquid flow must be readjusted and -- due to a short interruption of the liquid jet -- the gas flow control loop must return to a stable state.
        
        Figure~\ref{fig:combinedExp} also shows that, although the liquid flow is steady during normal operation, the measured absorption exhibits a large standard deviation as a result of the controller behavior.         However, averaging the results for the \cotwo[] flow rate over 
                the individual pseudo steady state periods yields values that follow a clear trend, as shown in the right panel of Figure~\ref{fig:combinedExp}. The representation in this panel is motivated by the model of the mass transfer in the LJA, which is presented below. 

        \subsubsection{Modeling} \label{subsubsec:LJA_modeling}

            The experimental results are evaluated assuming a cylindrical jet shape (jet radius $r_{0}$) and plug flow in the jet (axial flow velocity $u_{0} = \Dot{V} / (\pi r_{0}^{2})$). These are common assumptions for evaluating LJA experiments\cite{cullen_absorption_1957, danckwerts_gas-liquid_1970, johnson_diffusivity_1996, ying_measurements_2012} and discussed in more detail below.
    
\begin{figure}
                \centering
                \includegraphics[width=0.3\textwidth]{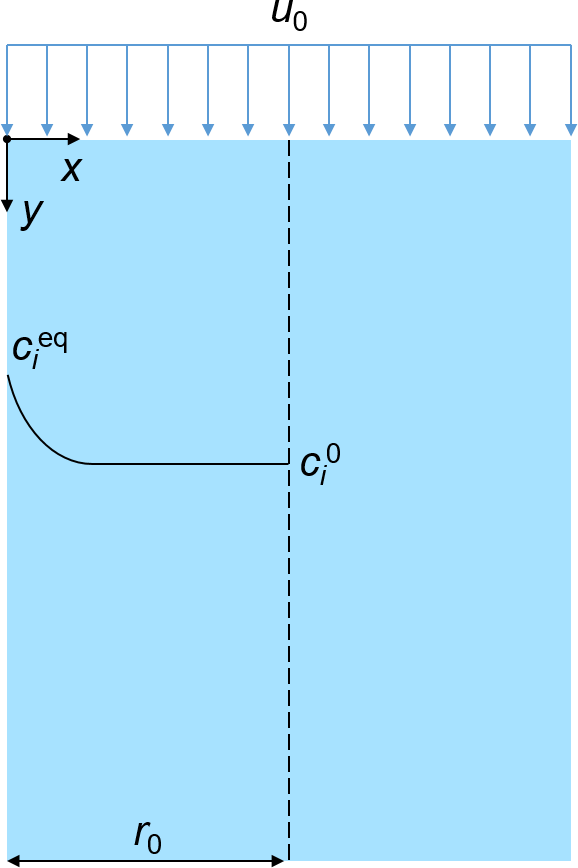}
                \caption{Sketch of the concentration profile for $t>0$ in the jet. The jet centerline (dashed line), initial velocity profile, coordinate axes, and the jet radius are also indicated. Figure not to scale. The liquid is assumed to be in equilibrium with the gas at the surface ($x=0$), leading to the molarity $c_{i}^{\mathrm{eq}}$, and is unloaded at large distances from the surface ($c_{i}^{0}=0$).}
                \label{fig:problemSketch}
            \end{figure}

            The problem is considered from a Lagrangian perspective, i.e., following a fluid particle along its trajectory in the jet. Furthermore, we assume that the concentration of the diffusing component $i$ (here: $i = \cotwo[]$) is small. Then, the transport of component $i$ in the mixture is purely diffusive, so that we have: 
            \begin{equation}
                \frac{\partial c_{i}}{\partial t} = \DIJ{i} \mathbf{\nabla}^{2} c_{i} \,,
                \label{eqn:diffusionEq}
            \end{equation}
            where $c_{i}$ is the molarity of component $i$ and \DIJ{i} is the Fick diffusion coefficient of the diluted component $i$ in the solvent. 
                                                                                    The problem is further simplified by neglecting axial diffusion and evaluating the resulting 1-D diffusion problem in a cartesian coordinate system, which gives practically the same results here, because diffusion -- due to the short contact times -- only takes place near the surface and the influence of the curvature is small (see \supp{} of Ref.~\cite{grossmann_hyrodynamics}).
                        Equation~\ref{eqn:diffusionEq} then becomes
            \begin{equation}
                \frac{\partial c_{i}}{\partial t} = \DIJ{i} \frac{\partial^{2}c_{i}}{\partial x^{2}} \,.
                \label{eqn:diffusionEqCartesian1D}
            \end{equation}      
            The diffusion problem is illustrated in Figure~\ref{fig:problemSketch}. Therein, $u_{0}$ is the axial velocity, which defines the relationship between the axial coordinate $y$ and the time $t$
            \begin{equation}
                y = u_{0} t \,.                 \label{eqn:axialTemporalRelationship}
            \end{equation} 
            
            The initial condition for the parabolic partial differential equation~(\ref{eqn:diffusionEqCartesian1D}) is given by the concentration of the diffusing component in the entry profile. 
            In the experiments of the present work, the entering liquid was unloaded, i.e., 
            \begin{equation}
                c_{i}(t=0) = c_{i}^{0} = 0 \,.
                \label{eqn:initialCondition}
            \end{equation}  
    
                        The concentration at the jet surface is constant and determined by the vapor-liquid equilibrium 
            \begin{equation}
                c_{i}(x=0) = c_{i}^{\mathrm{eq}} \,.
                \label{eqn:boundaryCondition}
            \end{equation}    
            In the present work, the equilibrium concentration $c_{i}^{\mathrm{eq}}$ was determined using the PCP-SAFT equation of state as described below. 

                        The contact times between the gas and the liquid are small in LJA experiments, so the jet can be considered as a semi-infinite slab, i.e., the concentration far from the surface (at large values of $x$) is the initial concentration $c_{i}^0$, which is zero here:
            
                        \changed{
            \begin{equation}
                                c(x \to {\infty}) = c_{0} = 0 \,,
                \label{eqn:boundaryCondition2}
            \end{equation} 
            }
                
            The solution of the present diffusion problem             is \cite{crank_mathematics_1975}
            \begin{equation}
                c_{i} = c_{i}^{\mathrm{eq}} \erfc{\left( \frac{x}{2\sqrt{\DIJ{i}t}} \right)} \,.
                \label{eqn:diffusionSolution}
            \end{equation}

                        From this, the molar flux of component $i$ is obtained from
            \begin{equation}
                j_{i} = - \DIJ{i} \frac{\partial c_{i}}{\partial x} \,.
                \label{eqn:fickLaw1}
            \end{equation}            
            yielding for the molar flux at the jet surface
            \begin{equation}
                j_{i}(x=0) = \sqrt{\frac{\DIJ{i}}{\pi t}}c_{i}^{\mathrm{eq}} \,.
                \label{eqn:molarDiffusiveFlow}
            \end{equation}        
                                                                        Integrating over time from $t=0$ up to the contact time $t=\tau$ yields an expression for the molar flow rate of component $i$, $\Dot{n}_{i}$, absorbed by the jet passing through the jet chamber:
            \begin{equation}
                \Dot{n}_{i} = \int_{t=0}^{\tau} j_{i}(x=0) 2 \pi r_{0} u_{0} \dd t \,,
                \label{eqn:NdotFromj}
            \end{equation} 
            where $r_{0}$ is the jet radius and $u_{0}$ the axial velocity, both of which are assumed to be constant. Inserting Eq.~\ref{eqn:molarDiffusiveFlow} into Eq.~\ref{eqn:NdotFromj} and introducing the jet length $L = u_{0} \tau$ yields
            \begin{equation}
                \Dot{n}_{i} = 4 c_{i}^{\mathrm{eq}} \sqrt{ \Dot{V} \DIJ[]{i} L } \,.
                \label{eqn:evaluationEquationNdot}
            \end{equation}         
            Eq.~(\ref{eqn:evaluationEquationNdot}) indicates a linear relationship between $\Dot{n}_{i}^{2}$ and $(4c_{i}^{\mathrm{eq}})^{2} \Dot{V} L$ with slope \DIJ{i}, which was confirmed in all experiments from the present work, see example in Figure~\ref{fig:combinedExp}. However, as can also be seen in Figure~\ref{fig:combinedExp}, a small offset from the origin is found, which can be attributed to inlet/outlet effects and does not compromise the result for \DIJ{i}.
            
            On average, about 10 data points $\Dot{n}_{i} \left( L \right)$ were acquired in each experiment. For all conditions, a repeat experiment was carried out on a different day; the results showed excellent repeatability, see Figure~\ref{fig:combinedExp}. All measurement points (i.e., both the original and the repeat experiment) were used in the linear regression of $\Dot{n}_{i}^{2}$ on $(4c_{i}^{\mathrm{eq}})^{2} \Dot{V} L$ to obtain a value for \DIJ{i}. 
            York regression \cite{york_unified_2004} was used, which is a special case of least squares that accounts for errors in the predictor and response variables. The uncertainties in the measured variables ($\Dot{n}_{i}, \Dot{V}, L$) were already specified above. The uncertainty in $c_{i}^{\mathrm{eq}}$ is estimated to be 5\%.             The             uncertainty in the diffusion coefficient \DIJ{i} is given by the 95\% confidence bounds of the corresponding regression coefficient and reported here individually for each experiment. Typical values for the resulting uncertainty of \DIJ{i} are about 10\%, which is distinctly higher than the deviations found in the repeat experiments that were only about 3\%.

        \subsection{PFG-NMR Spectroscopy}

            \subsubsection{Chemicals}
                1-Butanol (1-BuOH, $\ge$ 0.995 g g\textsuperscript{-1}) was purchased from Merck KGaA. For the PFG-NMR experiments, Carbon-\textsuperscript{13}C\,dioxide (\cotwo[13], $\ge$ 98.5\% enriched \cotwo[13]) was used, which was purchased from Sigma-Aldrich. The chemicals were used as received. 
                To compare the PFG-NMR results for \cotwo[13] with those for \cotwo[12] from the LJA, the former were corrected as described below.
                For simplicity, \cotwo[13] is simply labeled here as \cotwo[] wherever the differences are not important.     
            \subsubsection{\cotwo[] Gas Loading}
                The gas loading of \cotwo[] was carried out as in our previous works \cite{Behrens2017, Bellaire2022}. 1-BuOH (about 0.1~g) was filled into a high pressure valved NMR sample tube (Norell S-5-500-HW-EX1-HPV-7) and connected to a PTFE capillary, which was connected to a vacuum pump (Pfeiffer Vacuum PASCAL 2005SD) and to a \cotwo[] cylinder.                 The filled NMR sample tube was immersed in liquid nitrogen so that 1-BuOH freezes. After the sample was completely frozen, the gas over the frozen sample was removed using the vacuum pump until an absolute pressure of $p < 10^{-2}$~mbar was reached. 
                After this, the sample was brought back to room temperature.                 Then, the 1-BuOH was loaded with \cotwo[] at the desired pressure and temperature. For this, the sample tube was connected to a \cotwo[] cylinder equipped with a pressure reducing valve.
                                The pressure was measured with a calibrated pressure sensor (WIKA P-30). 
                The NMR sample tube was thermostatted in a water bath for 168~h (temperature control by a cryostat, Julabo F25-HE).

            \subsubsection{NMR Measurements}
            
                The measurements were carried out as in our previous work \cite{Bellaire2022}. \textsuperscript{13}C PFG-NMR measurements were carried out with an NMR spectrometer with a magnetic field strength of 9.4~T, corresponding to a proton Larmor frequency of 400.25~MHz, which was equipped with a double resonance broad band probe (magnet Ascend 400, console Avance III HD 400, probe BBFO, Bruker Biospin).                 The spectrometer’s temperature sensor was calibrated with a platinum resistance thermometer (Pt-100) that was, in turn, calibrated in our laboratory using a certified standard. The uncertainty of the temperature measurement is 0.1~K.

                After the gas loading, the NMR sample tube was placed inside the NMR spectrometer at the desired temperature. \textsuperscript{13}C PFG-NMR measurements were carried out using a pulse sequence with bipolar gradients (\texttt{stebpgp1s}) and \textsuperscript{13}C inverse gated \textsuperscript{1}H decoupled with 120~s relaxation delay. Each diffusion experiment consisted of 16 gradient steps with 16 scans. The gradient pulse duration $\delta$ was 2~ms. The gradient strength \textit{g} was incremented following a square root relationship from 2.3 to 43.1~G~cm\textsuperscript{-1} and the diffusion time $\mathit\Delta$ was chosen as 50~ms for all measurements. The time constant $\tau$ was chosen as 0.2~ms, corresponding to the time delay between the bipolar gradient lobes. Baseline and phase correction of the spectra were performed manually with MestReNova (Mestrelab Research). All peaks were integrated manually.	
        
                The self-diffusion coefficient $D_{\mathrm{CO}_2}$ was obtained from a fit of the natural logarithm of the relative NMR signal integral as a function of the gradient strength using the Stejskal-Tanner\cite{Stejskal1965} equation:
                \begin{equation}
                        \ln{\left(\frac{I}{I_{0}}\right)} = -D_{i}\gamma^{2}\delta^{2}g^{2}\left( \Delta-\frac{\delta}{3}-\frac{\tau}{2} \right)  \,,
                        \label{eqn:stejskalTanner}
                    \end{equation}
                where $I$ is the signal integral, $I_0$ is the signal integral without gradient and $\gamma$ is the gyromagnetic ratio of the \textsuperscript{13}C nuclei. The measurement was performed using \textsuperscript{13}C enriched \cotwo[] and therefore the self-diffusion coefficient was corrected, as in our previous work \cite{Bellaire2022}, with respect to the \textsuperscript{12}C isotope according to the following equation:
                    \begin{equation}
                        D_{^{12}\mathrm{CO}_{2}} = D_{^{13}\mathrm{CO}_{2}} \sqrt{\frac{M_{^{13}\mathrm{CO}_{2}}}{M_{^{12}\mathrm{CO}_{2}}}} \,,
                        \label{eqn:co2Correction}
                    \end{equation}  
                where $D_{i}$ are the self-diffusion coefficients and $M_{i}$ are the molar masses of the CO$_{2}$ isotopes. We present only the converted results for \DIJ[]{\cotwo[12]} here, the primary experimental PFG-NMR data is included in the \supp{}.
                
                To assess the measurement uncertainty, each experiment was repeated three times.                 The standard deviation of the three measurements was below $0.06 \times 10^{-9}~\mathrm{m}^2/\mathrm{s}$ in all cases. 

                               Diffusion coefficients at infinite dilution \DINF[]{\cotwo[12]} were obtained by extrapolation of the results from four experiments carried out with mixtures at low \cotwo[] concentrations (obtained by pressurization to 1, 2, 3, and 4 bar) at each temperature.                 In most cases, the results agreed within the experimental uncertainty, so that averaging was sufficient (see \supp{}). The uncertainty in the diffusion coefficient at infinite dilution is reported here individually for each experiment as the 95\% confidence interval for the extrapolated value \DINF[]{\cotwo[12]}.
            
        \subsection{PCP-SAFT EOS + DGT} \label{subsec:PCP-SAFT+DGT}
            
            The thermodynamic behavior of the mixture was modeled by the PCP-SAFT equation of state\cite{gross_perturbed-chain_2001, gross_application_2002, gross_equation--state_2005, gross_equation--state_2006, vrabec_vapor-liquid_2008}. 
                                                                                    
            For both \cotwo[] and 1-BuOH, the model parametrizations were taken from the literature and are listed in Table~\ref{tab:pureCompModels}. 
            
            \begin{table}
    \caption{Pure component parameters (segment number $m$, segment diameter $\sigma$, segment dispersion energy $\epsilon$, effective association volume $\kappa^{\mathrm{A}_{i}\mathrm{B}_{i}}$, association energy $\epsilon^{\mathrm{A}_{i}\mathrm{B}_{i}}$, dipole moment $D$, quadrupole moment $Q$) of the PCP-SAFT EOS for \cotwo[] \cite{gross_equation--state_2005} and 1-BuOH \cite{vins_density_2014}, as well as the DGT influence parameter $\kappa^{\mathrm{DGT}}$, taken from Ref.\cite{mairhofer_modeling_2017} for \cotwo[] and determined in the present work for 1-BuOH. $k_\mathrm{B}$ is the Boltzmann constant.}     
    \label{tab:pureCompModels}
    \begin{tabular}{ccccccccc}
        \hline
                & $m$       
                & $\sigma$  
                & $k_\mathrm{B} \epsilon$    
                & $10^{3} \times \kappa^{\mathrm{A}_{i}\mathrm{B}_{i}}$     
                & $k_\mathrm{B} \epsilon^{\mathrm{A}_{i}\mathrm{B}_{i}}$     
                & $D$     
                & $Q$   
                & $10^{20}\times\kappa^{\mathrm{DGT}}$                      
            \\
                & -         
                & Å         
                & K 
                & - 
                & K 
                & D    
                & DÅ 
                & Jm$^5$                    
            \\
        \hline
                \cotwo[]    
                & 1.5131    
                & 3.1869    
                & 163.33        
                & -         
                & -         
                & - 
                & 4.4   
                & 2.4197                    
            \\             
                1-BuOH 
                & 2.3838    
                & 3.7933    
                & 275.5        
                & 4.6   
                & 2678.0    
                & 1.66 
                & -     
                & 13.676 
            \\              
        \hline
    \end{tabular}
\end{table}

%
%
%
%

            As an alternative to the 1-BuOH model from \citet{vins_density_2014}, we have also tested the model of \citet{gross_application_2002}, which gave similar results that we do not report here for brevity.

            In the mixture ($i+j$), the binary dispersive-repulsive interactions are modeled by the modified Lorentz-Berthelot\cite{lorentz_ueber_1881,berthelot_sur_1898} combination rules
            \begin{equation}
                \sigma_{ij} = \frac{\sigma_{ij}+\sigma_{jj}}{2} \,,
                \label{eqn:LorentzBerthelotSigma}
            \end{equation}
            \begin{equation}
                \epsilon_{ij} = (1-k_{ij}) \sqrt{\epsilon_{ii}\epsilon_{jj}} \,,
                \label{eqn:LorentzBerthelotEpsilon}
            \end{equation}            
            with the binary interaction parameter $k_{ij}$. In this work,             the value of $k_{ij}$ in the mixture \cotwo[] + 1-BuOH was adjusted to experimental VLE data from the literature as described in the \supp{}, resulting in $k_{ij}=0.028$.

            The vapor-liquid interface is modeled by the density gradient theory (DGT), as described in our previous work\cite{grossmann_vapor-liquid_2024}.                                                                         The value of the pure component influence parameter for \cotwo[] was taken from the literature\cite{mairhofer_modeling_2017}, for 1-BuOH it was adjusted to literature data as described in the \supp{}.             The numerical values of the parameters are reported in             Table~\ref{tab:pureCompModels}.
            The influence parameter $\kappa_{ij}$ of the mixture $i+j$ was calculated from the pure-component influence parameters ($\kappa_{ii}$, $\kappa_{jj}$) using the geometric combination rule (Eq.~\ref{eq:dgt_kappaij}), without introducing an additional binary interaction parameter:
            \begin{equation}
                \kappa_{ij}  =  \sqrt{\kappa_{ii}\kappa_{jj}}  \,.
                \label{eq:dgt_kappaij}
            \end{equation}
            
            The density profiles $\boldsymbol{\rho}(z)$ in the interfacial region and the surface tension $\gamma$ were calculated as a function of the spatial coordinate $z$ using DGT in combination with the PCP-SAFT equation of state as described previously\cite{grossmann_vapor-liquid_2024}.

            Component density profiles may exhibit a local maximum in the interfacial region \cite{becker_interfacial_2016, stephan_molecular_2020, grossmann_vapor-liquid_2024}. In a previous work\cite{becker_interfacial_2016}, we have introduced the so-called enrichment $E_i$ to quantify this local maximum: $E_i$ is defined as the ratio of the highest local density of component $i$ to the larger of its two bulk phase component densities
            \begin{equation}
                E_{i} = \frac{\max (\rho_{i} (z))}{\max( \rho'_{i}, \rho''_{i})},
                \label{eq:DefEnrich}
            \end{equation}
            where the phase indices $'$ and $''$ denote the liquid and vapor phase, respectively.             
            From the density profiles in the interfacial region, also the relative adsorption and the interfacial thickness can be obtained as described in the \supp{}. Numerical data on the enrichment, relative adsorption, and interfacial thickness is provided in the \supp{} in machine-readable, tabular form.
        
    \section{Results and Discussion} \label{sec:resultsDiscussion}
               						   
    \subsection{Numerical Data}    The numerical results for the diffusion coefficients \DIJ[]{\cotwo[]} obtained from the evaluation of the LJA experiments are shown in Table~\ref{tab:table_results_LJA}. All experiments were conducted twice. As described in the experimental section, the repeatability was excellent. The data in Table~~\ref{tab:table_results_LJA} were obtained from a regression, in which the measurements from both experiments were used. 
    Details are given in the \supp{}, where we report the results for the molar flow rate $\dot{n}_{\cotwo[]}$ for all jet lengths $L$ from each individual experiment.
    The results for the diffusion coefficients \DIJ[]{\cotwo[]} at infinite dilution obtained from PFG-NMR are shown in Table~\ref{tab:table_resultsNMR}. Results for the individual experiments at finite concentrations are presented in the \supp{}.

    \begin{table}
    \caption{Diffusion coefficients of \cotwo[] in 1-BuOH obtained from the LJA experiments together with the estimated uncertainty.}
    \label{tab:table_results_LJA}
    \begin{tabular}{ccc}
        \hline
                $T$     
                & $p$
                & \DIJ[]{\cotwo[]}   
            \\        
                K 
                & bar
                & $10^{-9}\mathrm{m}^{2}\mathrm{s}^{-1}$
            \\
        \hline  
				283.63
				& 3.00
				& 1.50
				$\pm 0.19$
			\\
				283.93
				& 6.01
				& 1.57
				$\pm 0.28$
			\\
				283.84
				& 9.01
				& 1.72
				$\pm 0.20$
			\\
				284.07
				& 12.02
				& 1.86
				$\pm 0.22$
			\\
        \hline
				293.58
				& 6.01
				& 1.79
				$\pm 0.22$
			\\
				293.62
				& 9.01
				& 1.77
				$\pm 0.19$
			\\
				293.59
				& 12.02
				& 1.99
				$\pm 0.21$
			\\
        \hline
				303.18
				& 3.00
				& 1.82
				$\pm 0.23$
			\\
				303.28
				& 6.00
				& 1.95
				$\pm 0.20$
			\\
				303.29
				& 9.01
				& 1.99
				$\pm 0.20$
			\\
				303.32
				& 12.02
				& 2.35
				$\pm 0.25$
			\\
        \hline
				312.90
				& 6.00
				& 2.32
				$\pm 0.25$
			\\
				312.92
				& 9.01
				& 2.35
				$\pm 0.21$
			\\
				312.91
				& 12.01
				& 2.60
				$\pm 0.26$
			\\
        \hline
				322.90
				& 3.00
				& 2.24
				$\pm 0.29$
			\\
				322.94
				& 6.00
				& 2.65
				$\pm 0.25$
			\\
				322.95
				& 8.99
				& 2.93
				$\pm 0.18$
			\\
				322.95
				& 12.00
				& 3.22
				$\pm 0.30$
			\\
        \hline
				332.67
				& 6.00
				& 3.24
				$\pm 0.32$
			\\
				332.60
				& 9.00
				& 3.45
				$\pm 0.38$
			\\
				332.64
				& 12.00
				& 3.53
				$\pm 0.37$
			\\    
        \hline
    \end{tabular}
\end{table}

    \begin{table}
    \caption{Diffusion coefficients of \cotwo[] infinitely diluted in 1-BuOH obtained from PFG-NMR spectroscopy together with the estimated uncertainty.}
    \label{tab:table_resultsNMR}
    \begin{tabular}{cc}
        \hline
                $T$     
                & \DIJ[]{\cotwo[]}   
            \\        
                K 
                & $10^{-9}\mathrm{m}^{2}\mathrm{s}^{-1}$
            \\
        \hline  
                293.15
                & 2.66
                $\pm 0.03$
            \\            
                303.15
                & 3.09 
                $\pm 0.01$
            \\       
                313.15
                & 3.64    
                $\pm 0.01$
            \\ 
                323.15
                & 4.25  
                $\pm 0.02$
            \\             
        \hline
    \end{tabular}
\end{table}

    \subsection{Results from the Two Experimental Methods}     \label{subsec:ExpResults}

\begin{figure}
        \makebox[\textwidth][c]{
                        \includegraphics[width=0.8\textwidth]{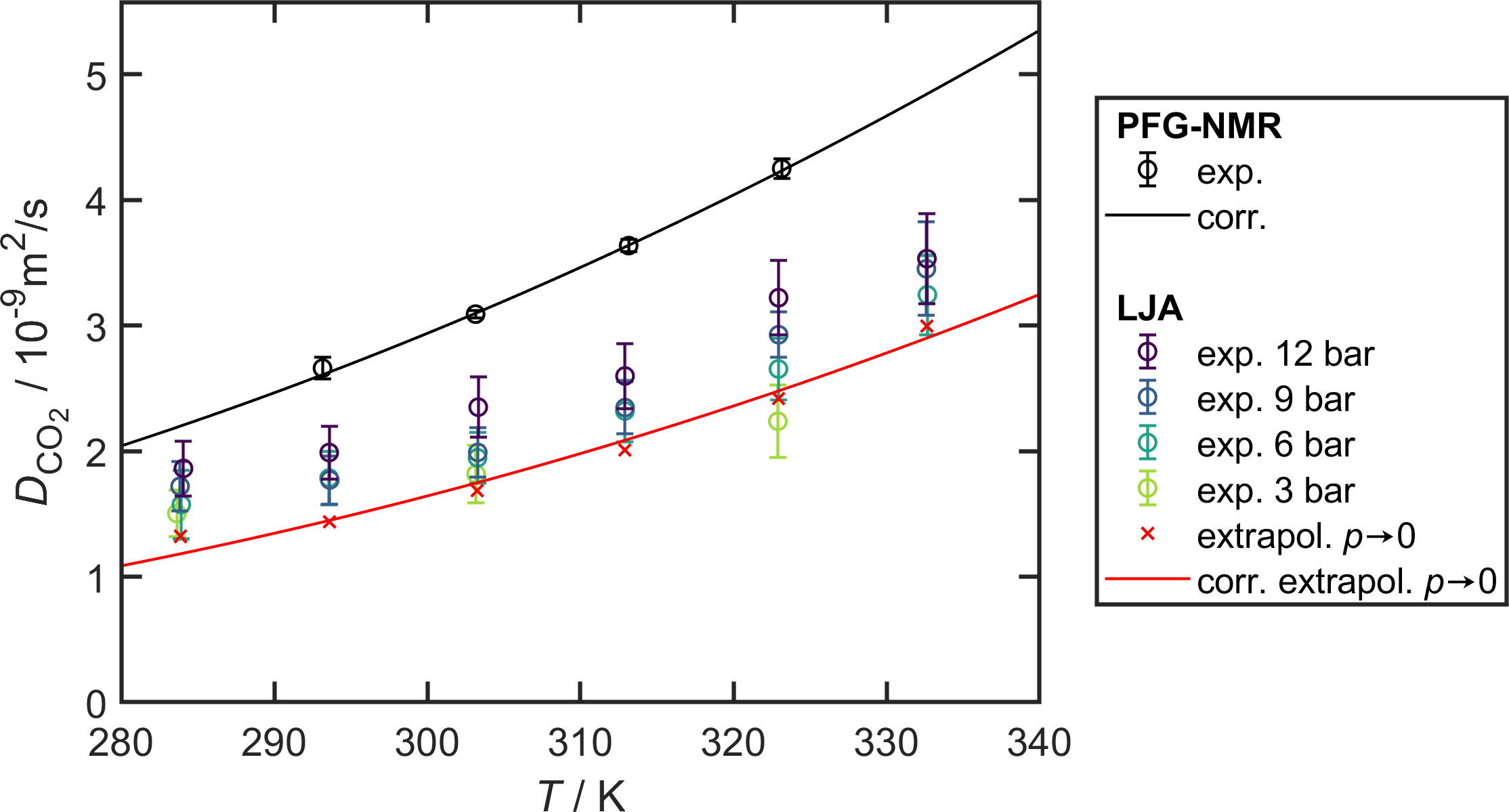}}
                    \caption{Overview of results for diffusion coefficients of \cotwo[12] in 1-BuOH obtained in the present work. Experimental results for infinite dilution obtained from PFG-NMR are compared to experimental data from LJA experiments carried out at different pressures, corresponding to different loadings of the liquid with \cotwo[]. To enable a direct comparison with the PFG-NMR data, the LJA data were extrapolated to low pressures, corresponding to infinite dilution of \cotwo[]. The lines are correlations of the data for infinite dilution.}         \label{fig:D_both}
    \end{figure}

    Figure~\ref{fig:D_both} compares the diffusion coefficients of \cotwo[] in 1-BuOH obtained using the two experimental methods employed in this study. 
    The PFG-NMR results for \cotwo[13] were converted into data for \cotwo[12] using Eq.~(\ref{eqn:co2Correction}) for the presentation in Figure~\ref{fig:D_both} and are data for infinite dilution of \cotwo[]. 
    The results obtained from the LJA are data for finite \cotwo[] concentrations and depend not only on temperature but also on pressure. The LJA data consistently show an increase of the results for \DIJ[]{\cotwo[]} with increasing pressure.
    This can have different reasons, \changed{which are discussed below.}

        \subsubsection{Concentration Differences}
    Even though the concentration profile of \cotwo[] in the laminar jet decays quickly with increasing distance from the jet's surface and the liquid jet is unloaded in its core, the diffusion takes place at finite concentrations of \cotwo[] with mole fractions of \cotwo[] near the surface of up to about 0.12~mol/mol,     see \supp{} and Figure~\ref{fig:mixVLE_Vins} below. 
        Increasing the pressure leads to increasing \cotwo[] concentrations near the surface and, hence, to an increase of the average \cotwo[] concentration in the region where diffusion takes place in the jet. 
                                            \changed{The concentration dependence of the LJA results shown in Figure~\ref{fig:D_both} could therefore simply result from the concentration dependence of the diffusion coefficient in the system \cotwo[] + 1-BuOH. 
    However, this dependence is expected to be weak in the concentration range studied in the present work. 
    First, for the related system \cotwo[] + 1-hexanol, experimental data for the Fick diffusion coefficient are available\cite{wu_mutual_2020} and show only a weak, unsystematic concentration dependence in this region. 
    Second, we carried out predictions for \cotwo[] + 1-BuOH using the hybrid machine-learning model EVE, which is based on the recently published ESE model for diffusion coefficients at infinite dilution\cite{Wagner2026} and enables prediction of the concentration dependence of the Fick diffusion coefficient. These predictions also indicate a weak concentration dependence for \cotwo[] + 1-BuOH, much too weak to explain the observed pressure dependence of the LJA data. Details are presented in the \supp{}.}
        
    \subsubsection{LJA Experiment Evaluation}
    Furthermore, we cannot exclude that the evaluation method used here for the evaluation of the LJA experiments with its simplifying assumptions might lead to an artificial pressure dependence of the data.
    However, the model assumptions are supported by the fact that the predicted linear dependence was always found in plots of the experimental results, like the one shown in Figure~\ref{fig:combinedExp}, right. Furthermore, the assumptions seem plausible for the present experiments and it is not evident how they could lead to an artificial pressure dependence of the results.

        The pressure dependence of the LJA data hampers a direct comparison with the PFG-NMR data. To nevertheless facilitate a direct comparison, the LJA data was simply extrapolated to low pressures, i.e., low loadings of the liquid with \cotwo[], as described in more detail in the \supp{}.
        
    Even though the extrapolation is subject to considerable uncertainties, it yields a constant trend. Despite all uncertainties, the overall message from Figure~\ref{fig:D_both} is clear: the LJA experiments yield substantially lower diffusion coefficients \DIJ[]{\cotwo[]} than the PFG-NMR experiments. These differences are analyzed in more detail below. Furthermore, Arrhenius-type correlations of the PFG-NMR data and the extrapolated data from the LJA were established, see Table~\ref{tab:table_DINF_correlation}, which are also included in Figure~\ref{fig:D_both}

    \begin{table}
    \caption{Correlations of the data for the diffusion coefficient of \cotwo[12] in 1-BuOH at infinite dilution from PFG-NMR and from LJA experiments:
    \\\hspace{\textwidth}
    \changed{$\DIJ[]{\cotwo[]} /(10^{-9}\mathrm{m}^{2}\mathrm{s}^{-1}) = A \cdot \exp \left( -B/(T/\mathrm{K}) \right)$}}     
    \label{tab:table_DINF_correlation}
    \begin{tabular}{ccc}
        \hline
                & $A$
                & $B$
            \\        
        \hline  
				PFG-NMR
				& 479.58
				& 1528.6
			\\
				LJA
				& 538.04 
				& 1737.5 
			\\
        \hline
    \end{tabular}
\end{table}

    \changed{\subsubsection{Comparison with Literature Data}
    Only three literature data points for diffusion coefficients in the system \cotwo[] + 1-BuOH are available, all at 298~K\cite{hikita_liquid_1959, takeuchi_simultaneous_1975}. We therefore carried out a broader literature study of experimental diffusion coefficients in systems for the type \cotwo[] + 1-alkanol. Data for the diffusion coefficient of \cotwo[] at infinite dilution are available for all 1-alkanols from methanol to 1-decanol, except 1-nonanol. The diffusion coefficient of \cotwo[] at infinite dilution decreases systematically with increasing chain length of the 1-alkanol. Comparison with this trend suggests that two of the three literature data points for \cotwo[] + 1-BuOH are too high. At the same time, the trend is consistent with our PFG-NMR results and indicates that the values obtained from the LJA are systematically too low. The literature data also support the temperature dependence observed in the present work. A detailed comparison is provided in the \supp{}.}

    \subsection{Extended Error Analysis} 
    A comparison of the results obtained with the two experimental methods shows that the diffusion coefficients measured by LJA are approximately 25--50\% lower than those from PFG-NMR. This discrepancy greatly exceeds the combined experimental uncertainties of both methods, as specified above, and therefore calls for an extended discussion of possible sources of error.

    \subsubsection{Jet Geometry and Flow}
    
    The evaluation of the LJA experiments is based on idealized assumptions, namely cylindrical jet geometry and an ideal plug flow. A detailed study of jets emerging from orifice plates like that used in the present work, which we have carried out by combining computational fluid dynamics and optical measurements\cite{grossmann_hyrodynamics}, shows that these assumptions are only approximations: There is a substantial jet contraction at the nozzle and a slight contraction further down the jet due to gravity. However, as explained above, the jet radius is not needed for the evaluation of the experiments, and most of the jet contraction takes place close to the nozzle, so that the assumption of cylindrical jet geometry seems acceptable. 
    The transition from the parabolic velocity profile in the feed line to a plug-flow profile in the free jet results in lower velocities near the jet surface than assumed using the plug flow assumption in the upper part of the jet.     However, the influence of this effect is masked at high Reynolds numbers as they were used in the present experiments\cite{grossmann_hyrodynamics}. This is also confirmed by the fact that no substantial deviations from the expected results were observed in the experiments carried out at different jet lengths.
    Hence, inlet effects and also outlet effects should not be a major source of error here. Furthermore, the plug flow assumption can only lead to an underestimation of the residence time and, therefore, to an overestimation of the diffusion coefficient.
    Thus, accounting for this effect, if it were to play a significant role, would only lead to an increase of the deviations between the PFG-NMR and the LJA results.

    Furthermore, we mention that all errors linked to the jet geometry and flow modeling depend only on the flow regime, characterized by the Reynolds, Froude, and Weber numbers, which, however, were identical during our experiments for a given temperature. Thus, the significant pressure dependence of the LJA results for the diffusion coefficient at a given temperature (cf. Figure~\ref{fig:D_both}) cannot be explained by such errors.

    \subsubsection{Jet Capture}
    
    If the laminar jet does not enter the capture funnel properly, splashing or gas entrainment may occur, especially during the experiment startup or for jet lengths approaching the breakup length, where surface waves of significant amplitude arise\cite{grossmann_hyrodynamics}. In such cases, larger absorption rates will be measured. Furthermore, splashed liquid remaining in the chamber could offer excess surface area for absorption, but only until it is saturated. In all cases, this would lead to measured values of the diffusion coefficient that are too high, and a correction would increase the deviations between the LJA results and those from PFG-NMR, see Figure~\ref{fig:D_both}.

    \subsubsection{Thermophysical Property Data}

    Applying Eq.~(\ref{eqn:boundaryCondition}) for the evaluation of the primary data from the LJA experiments to determine \DIJ[]{\cotwo[]} requires the \cotwo[] concentration at the jet surface $c^{\mathrm{eq}}_{\cotwo[]}$. 
    This concentration was obtained from a vapor-liquid equilibrium calculation with the PCP-SAFT EOS, specifying the experimental temperature and pressure. The value obtained for the diffusion coefficient \DIJ[]{\cotwo[]} is highly sensitive to $c^{\mathrm{eq}}_{\cotwo[]}$ as there is an inverse quadratic relationship between the two in Eq.~(\ref{eqn:evaluationEquationNdot}).
    However, as shown below, the relative uncertainty of the value for $c^{\mathrm{eq}}_{\cotwo[]}$ does not exceed 8\% even in the worst cases, which translates into 15\% in the diffusion coefficient, and is typically well below that. Thus, that uncertainty is not sufficient to explain the magnitude of the deviations between the LJA and PFG-NMR results.     
            
    \subsubsection{Gas Phase Resistance}

    The previously discussed relationship at the jet surface (c.f. Eq.~(\ref{eqn:boundaryCondition})) applies only when gas phase resistance can be neglected.     The flow pattern in the laminar jet chamber, in which the liquid jet induces a circulating flow in the gas phase, is briefly discussed in Ref.\cite{grossmann_hyrodynamics}.
                Based on the results from Ref.\cite{grossmann_hyrodynamics}, the influence of the gas phase resistance can be estimated using numerical simulations. We have carried out such simulations based on a scenario in which the bulk gas phase is coupled to the liquid phase, described by penetration theory, by a gas-side boundary layer for which the mass transport equations are solved.
    Details of that simulation are given in the \supp.
    These preliminary results confirm that, for all conditions relevant to our LJA experiments, the gas-side mass transfer resistance is significantly lower than on the liquid side. Hence, for the short contact times in the LJA (here: $\leq 32$~ms), $c^{\mathrm{eq}}_{\cotwo[]}$ is a good approximation for the surface concentration. 
    Deviations that could be attributed to the gas-side mass transfer are well below those from the uncertainty of the phase equilibrium calculation. Hence, the gas-side mass transfer resistance cannot explain the very low values for \DIJ[]{\cotwo[]} found in the LJA experiments and accounting for that resistance would lead to no substantial increase of the observed values of the diffusion coefficient.

    \subsubsection{Hypothesis}

        The results from PFG-NMR are considered as ground truth here, because the experiment is very simple, at least compared to the LJA experiment, and the diffusion process is monitored directly without any interference with the system. 
    The way to evaluate the PFG-NMR results is firmly established and the error analysis is straightforward.
    PFG-NMR is therefore commonly considered as a prime source for high quality data on diffusion coefficients.\cite{holz_temperature-dependent_2000, weingartner_nmr_2002, pages_pulsed-field_2017, Bellaire2022, steimers_accurate_2022, phuong_determination_2024}
        We therefore conclude that the problem must lie on the LJA side. 
    While there are many possible sources of error in the LJA experiment and its evaluation, their extended analysis has yielded no explanation for the large discrepancies to the results from PFG-NMR.
    The most plausible remaining hypothesis to explain the low values of \DIJ[]{\cotwo[]} observed in the LJA experiments is the presence of an additional mass transfer resistance at the interface. This hypothesis is examined in the following based on results obtained from DGT in combination with the PCP-SAFT EOS. Thereby, we build on a second hypothesis, postulating that the interfacial mass transfer resistance is related to the enrichment of \cotwo[] at the interface.

    \subsection{PCP-SAFT EOS + DGT} 
\begin{figure}
            \centering
            \includegraphics[width=0.6\textwidth]{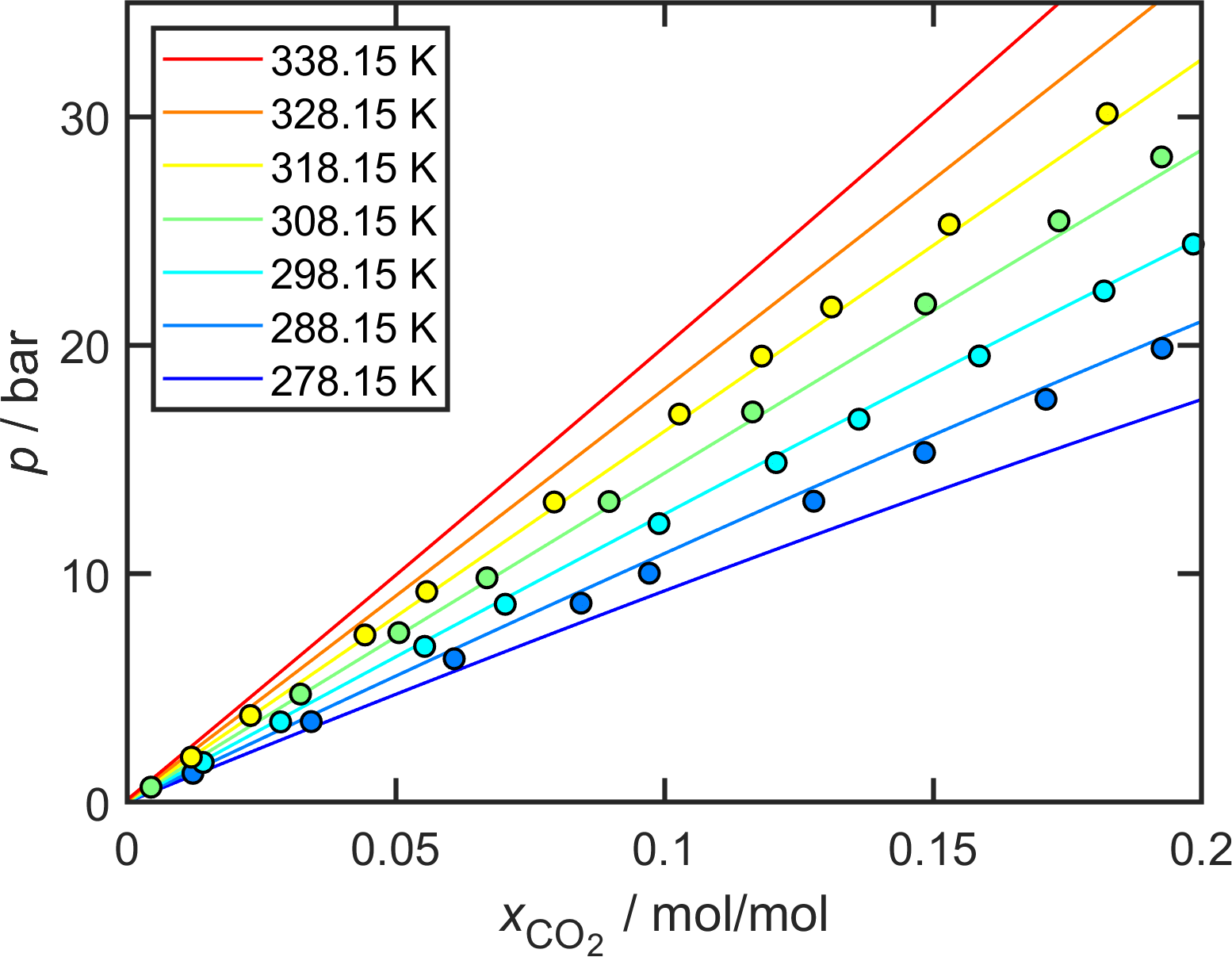}
            \caption{Vapor-liquid equilibrium of the mixture \cotwo[] + 1-BuOH at different temperatures: Experimental data from \citet{gui_solubility_2011} (symbols) are compared to the PCP-SAFT model from the present work (lines).}            \label{fig:mixVLE_Vins}
        \end{figure}

                The EOS model from the present work with the adjusted $k_{ij}$ describes the vapor-liquid equilibrium of the mixture \cotwo[] + 1-BuOH well, as can be seen from the comparison with experimental data of \citet{gui_solubility_2011} in Figure~\ref{fig:mixVLE_Vins}.
        The absolute error in the liquid phase mole fraction $x_{\cotwo[]}$ calculated from the temperature $T$ and the pressure $p$ does not exceed 0.006~mol/mol in the entire temperature and pressure range relevant for the present measurements, the mean absolute error is only 0.002~mol/mol. The PCP-SAFT prediction of three further isotherms between 278.15~K and 338.15~K, beyond the experimental data of \citet{gui_solubility_2011}, is also included in Figure~\ref{fig:mixVLE_Vins} to cover the temperature range relevant to the present experimental work.

\begin{figure}
            \centering
            \includegraphics[width=0.6\textwidth]{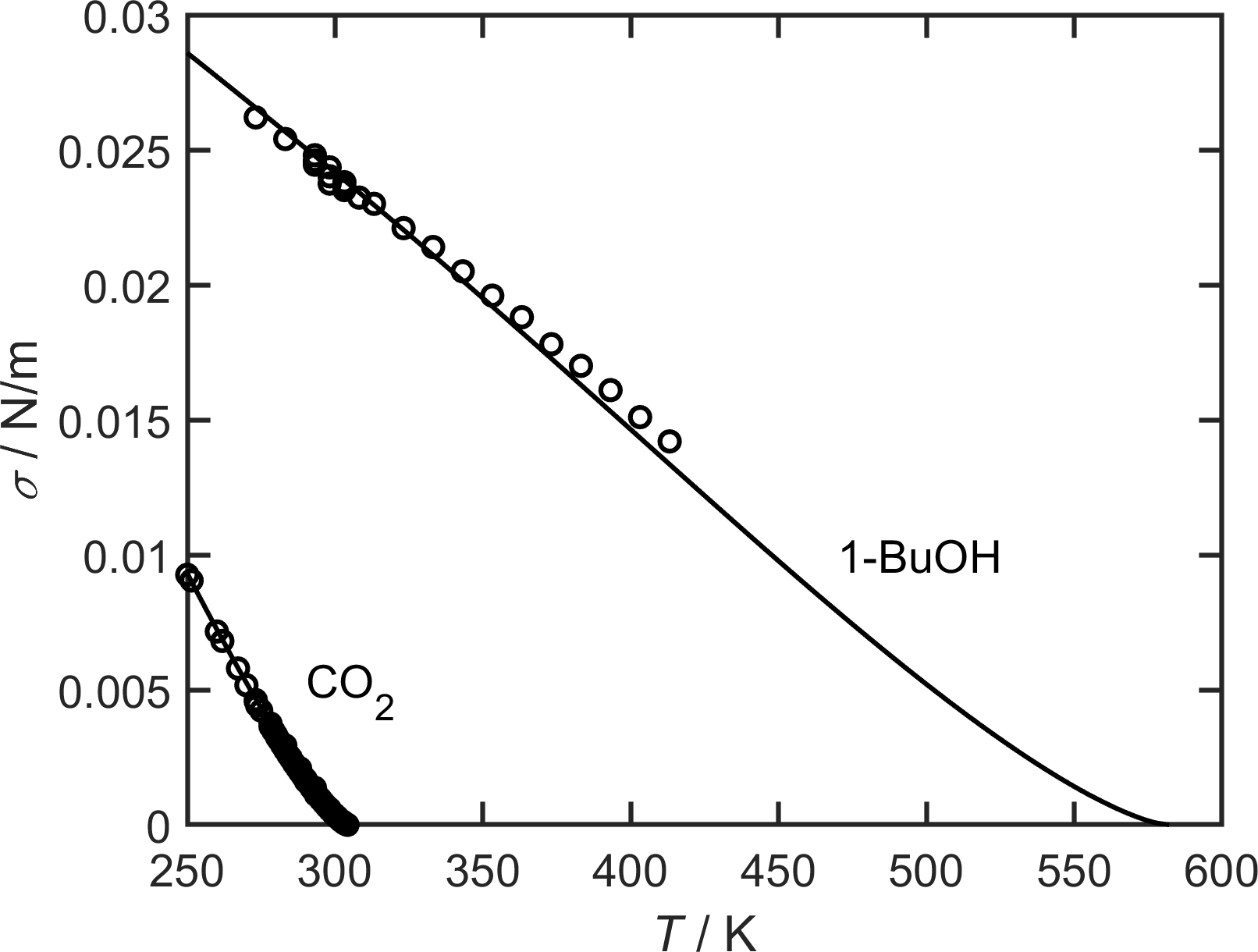}
            \caption{Pure component surface tensions for \cotwo[] and 1-BuOH: Experimental data from the literature\cite{mumford_19_1950, timmermans_physico-chemical_1965, riddick_organic_1970, blake_adsorption_1972, vargaftik_tables_1975, jimenez_excess_2001,quinn_surface_1927, hahne_heat_1977, muratov_surface_1982, pearce_light-scattering_1987} (symbols) are compared to the PCP-SAFT+DGT models from the present work (lines).}
            \label{fig:pureSFT}
        \end{figure}

                As shown in Figure~\ref{fig:pureSFT}, the combination of PCP-SAFT + DGT model from the present work describes experimental data of the surface tension of the pure components \cotwo[] and 1-BuOH well using the adjusted influence parameters, see Table~\ref{tab:pureCompModels}.         No binary data on the surface tension is available to test the model predictions.

        The PCP-SAFT + DGT model was used to predict density profiles in the vapor-liquid interfacial region of the mixture \cotwo[] + 1-BuOH for all conditions that were studied experimentally in the present work, from which data for the mixture surface tension, relative adsorption, enrichment, and interfacial thickness were calculated. These data are provided in tabular form in the \supp{}. 
        
\begin{figure}
            \centering
            \includegraphics[width=0.6\textwidth]{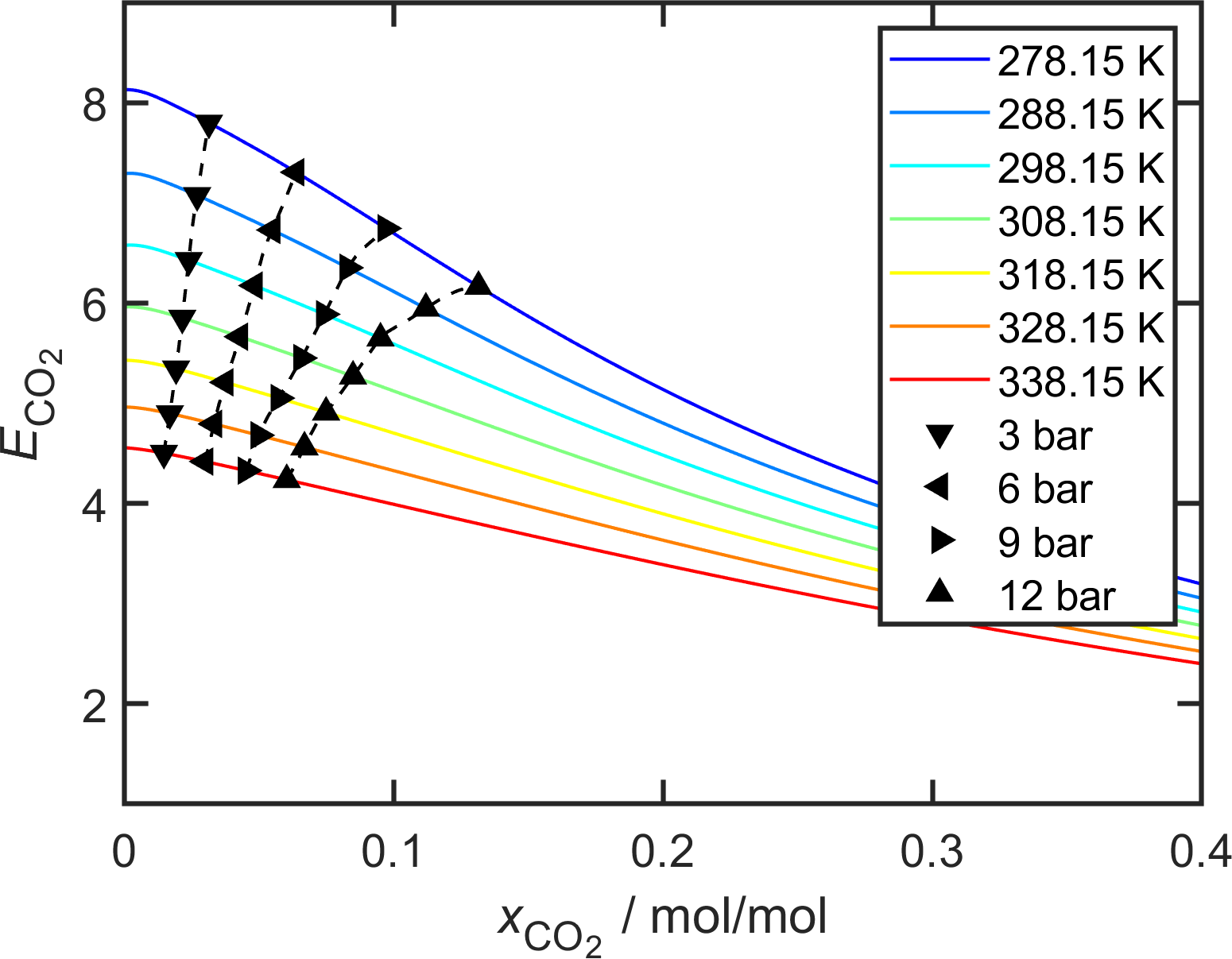}
            \caption{Enrichment $E_{\cotwo[]}$ of \cotwo[] at the vapor-liquid interface in mixtures of \cotwo[] and 1-BuOH as a function of the liquid phase mole fraction of \cotwo[], $x_{\cotwo[]}$, for different temperatures (colored lines). Symbols indicate the pressure. Results from the PCP-SAFT EOS combined with DGT.}
                                    \label{fig:Enrichments}
        \end{figure}

        For the following discussion, only the enrichment $E_{\cotwo[]}$ is relevant, which is plotted in Figure~\ref{fig:Enrichments} as a function of the liquid phase \cotwo[] mole fraction, $x_{\cotwo[]}$, for several isotherms that encompass the range studied in the present work.        
        For all conditions, the enrichment $E_{\cotwo[]}$ is between about 4 and 8, which are exceptionally high values\cite{stephan_enrichment_2020}, i.e., the concentration of \cotwo[] at the interface is several times higher than in either the bulk liquid or vapor phase. 
        The enrichment decreases with increasing temperature and increasing concentration of \cotwo[] in the liquid, i.e., with higher pressures.
        An exception occurs only at very low concentrations ($x_{\cotwo[]} < 0.01$~mol/mol), where a slight local maximum in $E_{\cotwo[]}$ is predicted for the lowest studied isotherms.         
        
\begin{figure}
            \centering
            \includegraphics[width=0.6\textwidth]{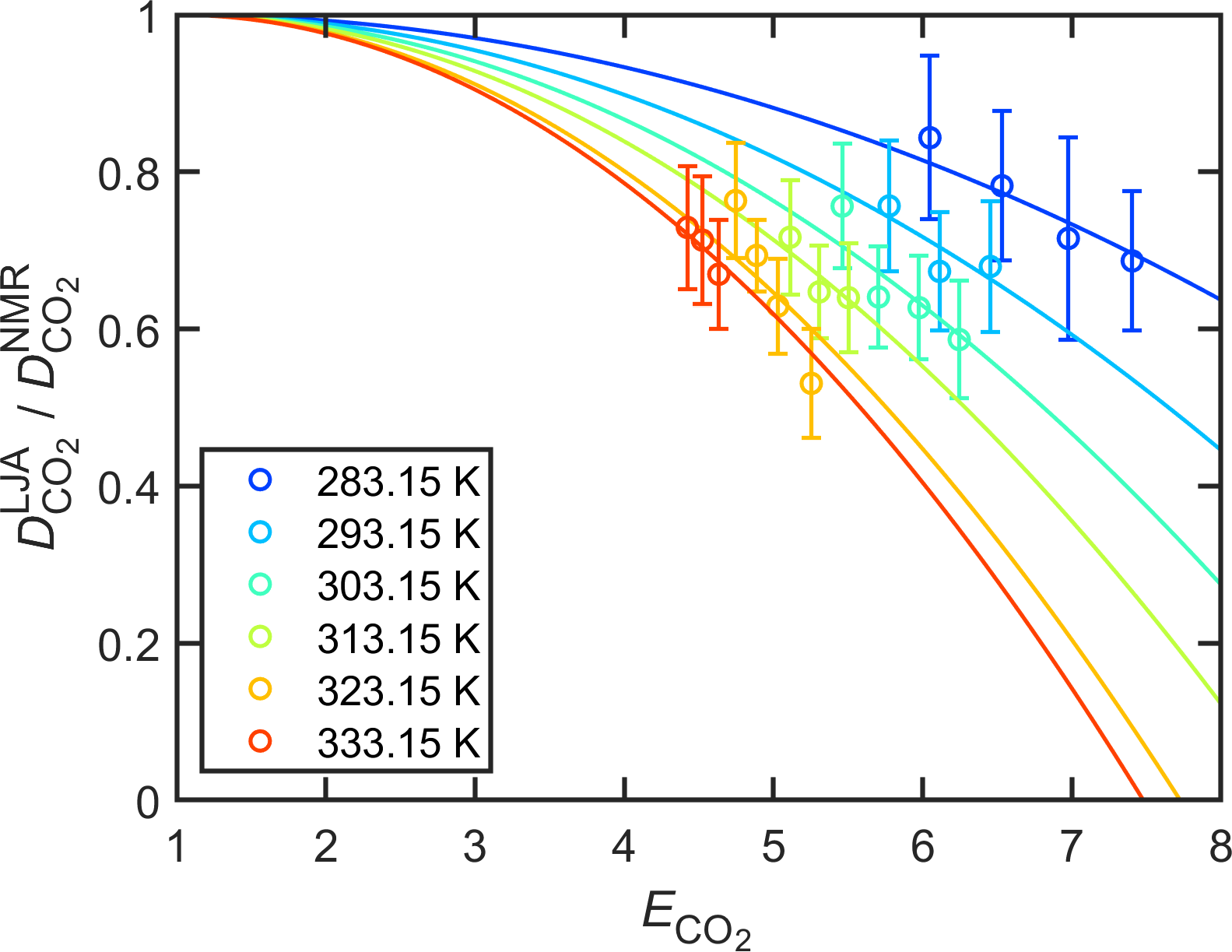}
            \caption{Ratio of \DIJ[]{\cotwo[]} from LJA to NMR measurements as a function of the enrichment $E_{\cotwo[]}$ for several temperatures. \changed{The errorbars stem from the propagated uncertainties of the LJA and NMR data.} Lines are quadratic fits to the data.}
            \label{fig:D_ratio}
        \end{figure}

        To investigate whether the enrichment is correlated to the deviation between the measurements from LJA and PFG-NMR, the ratio of the two diffusion coefficients of \cotwo[] obtained from the respective experiments (\DcotwoSup[LJA]{} / \DcotwoSup[NMR]{}) was plotted versus $E_{\cotwo[]}$ for all experimental conditions. 
        Where no experimental PFG-NMR data were available, the \DcotwoSup[NMR]{} values were taken from the correlation from Table~\ref{tab:table_DINF_correlation}.
        Figure~\ref{fig:D_ratio} shows the resulting plot. As guide to the eye, quadratic correlations of the isothermal data are shown that pass the origin ($E_{\cotwo[]} = 1, \DcotwoSup[LJA]{} / \DcotwoSup[NMR]{}=1$), implying that, if no enrichment is present, the deviations vanish.         
        
        Although Figure~\ref{fig:D_ratio} suggests a correlation between the enrichment $E_{\cotwo[]}$ and the deviation $\DcotwoSup[LJA]{} / \DcotwoSup[NMR]{}$, a close inspection reveals caveats.
        Most importantly, the deviations are fairly independent of temperature, whereas the enrichment is much higher at low temperatures. 
        As a consequence, one would have to postulate an important influence of the temperature on the dependence of the deviation $\DcotwoSup[LJA]{} / \DcotwoSup[NMR]{}$ on the enrichment $E_{\cotwo[]}$: At constant enrichment, the deviations decrease with increasing temperature, cf. Figure~\ref{fig:D_ratio}.
                
        Let us, with all due caution, for a moment assume that a causal relation between $\DcotwoSup[LJA]{} / \DcotwoSup[NMR]{}$ and $E_{\cotwo[]}$ exists and that it is quadratic for a given temperature, see Figure~\ref{fig:D_ratio}. 
        According to our hypothesis, $E_{\cotwo[]}$ causes a mass transfer resistance at the vapor-liquid interface. 
        To account for this in the evaluation of the LJA experiment (cf. Eq.~(\ref{eqn:evaluationEquationNdot})), the concentration of \cotwo[] at the jet surface would no longer be the equilibrium concentration $c_{\cotwo[]}^{\mathrm{eq}}$ but some concentration $c_{\cotwo[]}^{*}$ instead, with $c_{\cotwo[]}^{*} / c_{\cotwo[]}^{\mathrm{eq}} < 1$. 
        Let us furthermore assume that the PFG-NMR result is the ground truth and a perfect match with this result would be obtained using the value $c_{\cotwo[]}^{*}$ for the concentration at the jet surface. 
        According to Eq.~(\ref{eqn:evaluationEquationNdot}), the diffusion coefficient depends quadratically on the inverse of the concentration at the jet surface, so that we have:
        \begin{equation}
            \frac{\DcotwoSup[LJA]{}}{\DcotwoSup[NMR]{}} = 
            \left(
                \frac{c_{\cotwo[]}^{*}}{c_{\cotwo[]}^{\mathrm{eq}}}
            \right)^2            \,,
            \label{eqn:nmrCorrelation}
        \end{equation}
        where \DcotwoSup[LJA]{} and $c_{\cotwo[]}^{\mathrm{eq}}$ are taken from the unmodified evaluation of the LJA results. Eq.~(\ref{eqn:nmrCorrelation}) enables obtaining estimates for $c_{\cotwo[]}^{*}$. 
        Furthermore, Figure~\ref{fig:D_ratio} suggests that $\DcotwoSup[LJA]{} / \DcotwoSup[NMR]{}$ depends quadratically on $E_{\cotwo[]}$.
        Combining this with Eq.~(\ref{eqn:nmrCorrelation}) would imply a linear dependence of $c_{\cotwo[]}^{*} / c_{\cotwo[]}^{\mathrm{eq}}$ on $E_{\cotwo[]}$.

    \section{Conclusions}                                                         This study presents new experimental data on diffusion coefficients of carbon dioxide in binary mixtures with 1-butanol, for which no data were previously available. The measurements were carried out using two fundamentally different experimental techniques: a newly developed pressurized laminar jet apparatus (LJA) and pulsed field gradient NMR spectroscopy (PFG-NMR). Our novel LJA design was specifically engineered to enable measurements at elevated pressures (up to 12~bar) and considerably extends the operational window of this measurement technique. PFG-NMR measurements served as a reference for diffusion in the absence of interfacial mass transfer.
        Both methods provided precise and repeatable measurements across a wide temperature range. However, systematic deviations were observed between the two techniques: The LJA consistently yielded lower diffusion coefficients than PFG-NMR. An extensive analysis ruled out experimental error and uncertainties introduced by the model-based evaluation of the experimental data as primary sources of the observed discrepancy.

        Adopting two hypotheses from the literature, we assume for the further investigation that the discrepancies are caused by a mass transfer resistance at the vapor-liquid interface, which is present only in the LJA experiments, and that this resistance is caused by the enrichment of the diffusing species \cotwo[] at the interface.
        The enrichment of \cotwo[] in the studied system was calculated using the PCP-SAFT equation of state combined with density gradient theory and found to be exceptionally high. For the experimental conditions, the molarity of \cotwo[] in the interfacial region is 4--8 times higher than in the bulk liquid phase, which itself is higher than the molarity in the bulk gas phase.
        For a given temperature, the discrepancies between the results from the LJA experiments and those of the PFG-NMR experiments, which are considered as a ground truth here, were found to correlate well with the enrichment by a simple quadratic relationship. 
        This enables estimation of the deviation from interfacial equilibrium, expressed by the ratio ${c_{\cotwo[]}^{*}} / c_{\cotwo[]}^{\mathrm{eq}}$, which is found to depend linearly on the enrichment $E_{\cotwo[]}$.

                These findings are, however, still exploratory and highlight the need for further investigations, both experimental and theoretical, into the impact of nanoscale interfacial properties on macroscopic mass transfer in gas-liquid systems.
        This work establishes a feasible route for systematic studies of this important topic. At the same time, it contributes to availability of diffusion coefficient data and the development of reliable methods for their measurement.     
    \begin{acknowledgement}
        The authors gratefully acknowledge funding of the present work by the ERC Advanced Grant ENRICO (Grant Agreement No. 694807). The authors further thank Prof. Dr.-Ing. Dr. rer. nat. Simon Stephan of OVG University Magdeburg for access to the \textit{MicTherm} code package for the calculation of the EOS and DGT results.
    \end{acknowledgement}

    \begin{suppinfo}
                
        The following files are available free of charge.
        \begin{itemize}
          \item \texttt{SI.pdf}: Supplementary text to the main manuscript.
          \item \changed{\texttt{SI\_data.xlsx}: Two machine-readable tables in separate worksheets: 
          a) raw experimental data from the laminar jet apparatus and all parameters required to calculate the diffusion coefficient from the mass transfer, and 
          b) data calculated with density gradient theory in combination with the PCP-SAFT equation of state.}
        \end{itemize}
    
    \end{suppinfo}    
        
    \bibliography{LJA}

\end{document}